\documentclass{elsart3}

\usepackage{graphics}
\usepackage{graphicx}
\usepackage{epsfig}

\usepackage{amssymb}

\begin{document}

\begin{frontmatter}



\vspace{-2.cm}
\title{BAIKAL Experiment: main results obtained with the 
neutrino telescope NT200}


\author[a]{V. Aynutdinov},
\author[a]{V. Balkanov},
\author[d]{I. Belolaptikov},
\author[a]{L. Bezrukov},
\author[a]{D. Borschov},
\author[b]{N. Budnev},
\author[a]{I. Danilchenko},
\author[a]{Ya. Davidov},
\author[a]{G. Domogatsky},
\author[a]{A. Doroshenko},
\author[b]{A. Dyachok},
\author[a]{Zh.-A. Dzhilkibaev\corauthref{cor}},
\corauth[cor]{Corresponding author.}
\ead{djilkib@pcbai10.inr.ruhep.ru}
\author[f]{S. Fialkovsky},
\author[a]{O. Gaponenko},
\author[d]{K. Golubkov},
\author[b]{O. Gress},
\author[b]{T. Gress},
\author[b]{O. Grishin},
\author[a]{A. Klabukov},
\author[h]{A. Klimov},
\author[a]{S. Klimushin},
\author[b]{A. Kochanov},
\author[d]{K. Konischev},
\author[a]{A. Koshechkin},
\author[c]{L. Kuzmichev},
\author[f]{V. Kulepov},
\author[a]{B. Lubsandorzhiev},
\author[a]{S. Mikheyev},
\author[e]{T. Mikolajski},
\author[f]{M. Milenin},
\author[b]{R. Mirgazov},
\author[c]{E. Osipova},
\author[b]{A. Pavlov},
\author[b]{G. Pan'kov},
\author[b]{L. Pan'kov},
\author[a]{A. Panfilov},
\author[a]{D. Petukhov},
\author[d]{E. Pliskovsky},
\author[a]{P. Pokhil},
\author[a]{V. Poleschuk},
\author[c]{E. Popova},
\author[c]{V. Prosin},
\author[g]{M. Rozanov},
\author[b]{V. Rubtzov},
\author[b]{Yu. Semeney},
\author[a]{B. Shaibonov},
\author[c]{A. Shirokov},
\author[e]{Ch. Spiering},
\author[b]{B. Tarashansky},
\author[d]{R. Vasiliev},
\author[e]{R. Wischnewski},
\author[c]{I. Yashin},
\author[a]{V. Zhukov}

\address[a]{Institute for Nuclear Research, 60th October Anniversary pr. 7a, 
Moscow 117312, Russia}
\address[b]{Irkutsk State University, Irkutsk, Russia}
\address[c]{Skobeltsyn Institute of Nuclear Physics  MSU, Moscow, Russia}
\address[d]{Joint Institute for Nuclear Research, Dubna, Russia}
\address[e]{DESY, Zeuthen, Germany}
\address[f]{Nizhni Novgorod State Technical University, Nizhni Novgorod, 
Russia}
\address[g]{St.Petersburg State Marine University, St.Petersburg, Russia}
\address[h]{Kurchatov Institute, Moscow, Russia}

\begin{abstract}
The Baikal Neutrino Telescope  NT200 takes data since April 1998.
	On April 9th, 2005, the 10 Mton scale detector NT200$+$ was put into
	operation in Lake Baikal. 
	Selected results obtained during 1998-2002 with the neutrino telescope
	NT200 are presented.
\end{abstract}

\begin{keyword}
Neutrino telescopes \sep Neutrino astronomy \sep UHE neutrinos \sep BAIKAL

\PACS 95.55.Vj \sep 95.85.Ry \sep 96.40.Tv
\end{keyword}
\end{frontmatter}

\section{Introduction}
The Baikal Neutrino Telescope  NT200 takes data since April 1998.
On April 9th, 2005, the 10-Mton scale detector NT200$+$ was put into
operation in Lake Baikal. 
Description of site properties, detector configuration and
performance have been described elsewhere \cite{NE05,HEB,NANP05,RW}.

In this paper we present main results obtained from the analysis of data
taken with the Baikal neutrino telescope NT200 between April 1998 
and February 2003. 

\section{Selected results obtained with NT200}
\vspace{-0.3cm}
\subsection{Atmospheric neutrinos}
\begin{figure}[htb]
\vspace{-1.0cm}
\includegraphics[width=7.5cm,height=7.5cm]{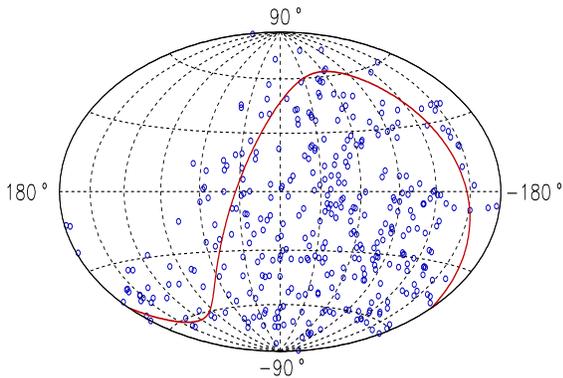}
\vspace{-1.0cm}
\caption{
  Skyplot (galactic coordinates) of neutrino events for five years. 
  The solid curve shows the equator.
}
\label{fig1}
\end{figure}

\vspace{-0.3cm}
The signature of charged current muon neutrino events is a muon 
crossing the detector from below. Muon track reconstruction
algorithms and background rejection have been described elsewhere \cite{APP2}.
Compared to \cite{APP2} the analysis of the 4-year sample (1038 days
live time) was optimized for higher signal passing
rate, and accepting a slightly higher contamination of 15-20\%
fake events \cite{ECRS04}.
A total of 372 upward 
going neutrino candidates were selected.
From Monte-Carlo simulation a total of 385 atmospheric neutrino
and background events are expected.
The skyplot of these events is shown in Fig. \ref{fig1}.

\vspace{-0.3cm}
\subsection{Search for Neutrinos from WIMP Annihilation}
\vspace{-0.3cm}
The search for WIMPs with the Baikal
neutrino telescope is based on a possible signal of
nearly vertically upward going muons, exceeding
the flux of atmospheric neutrinos. 
The method of event selection relies on 
the application of a series of cuts which are tailored to the response
of the telescope to nearly vertically upward moving muons 
\cite{Bal1}.
The applied cuts select muons with -1$< \cos(\theta) <$-0.75 
and result in a detection area of about 1800 m$^2$ for vertically 
upward going muons.
\begin{figure}[htbp]
\includegraphics[width=.4\textwidth]{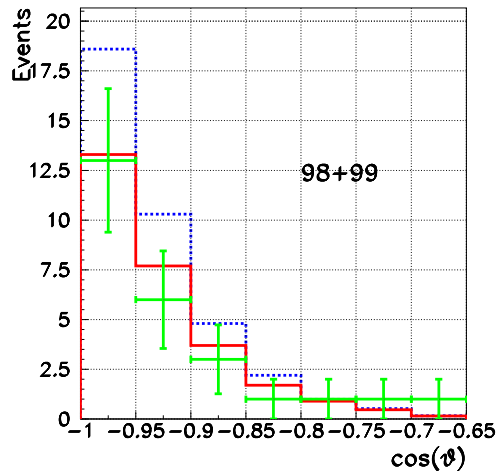}
 \caption{Angular distributions of selected neutrino
candidates as well as expected distributions in a case
with and without oscillations (solid and dashed curves
respectively).
}
\label{fig2}
\end{figure}
The energy threshold for this analysis is $E_{\mbox{thr}}\sim 10$ GeV
i.e. significantly lower then for the
analysis described in \mbox{section 2.1} ($E_{\mbox{thr}}\sim 15 - 20$ GeV).
Therefore the effect of oscillations is stronger visible.
We expect a muon event suppression of (25-30)\% due to neutrino oscillations
assuming 
$\delta m^2 = $2.5$\cdot$10$^{-3}$ eV$^2$ 
with full mixing, $\theta_m\approx \pi/$4.

From 502 days of effective data taking 
between April 1998 and February 2000,
24 events with -1$< \cos(\theta) <$-0.75 have been selected 
as clear neutrino events.
The angular distribution of these events
as well as the MC - predicted distributions 
are shown in \mbox{Fig. \ref{fig2}}. 
For the MC simulations we used the Bartol96 atmospheric
neutrino flux \cite{Bartol}
without (dashed curve) and with (solid curve) oscillations. 
Within 1$\sigma$ statistical uncertainties the experimental
angular distribution is  consistent with the prediction
including neutrino oscillations.

\begin{figure}[htbp]
\includegraphics[width=.40\textwidth,height=6.cm]{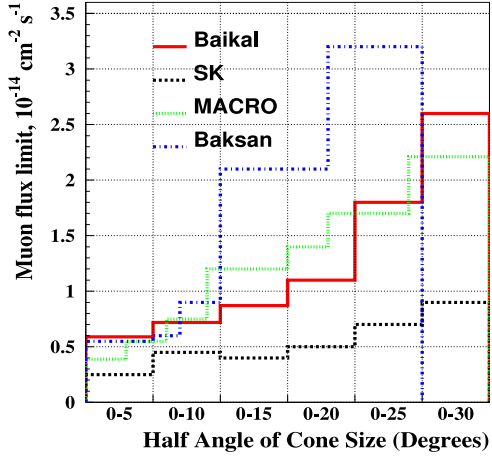}
 \caption{Limits on the excess muon flux from the center of 
the Earth versus half-cone of the search angle. 
}
\label{fig3}
\end{figure}

\begin{figure}[htbp]
\includegraphics[width=.45\textwidth]{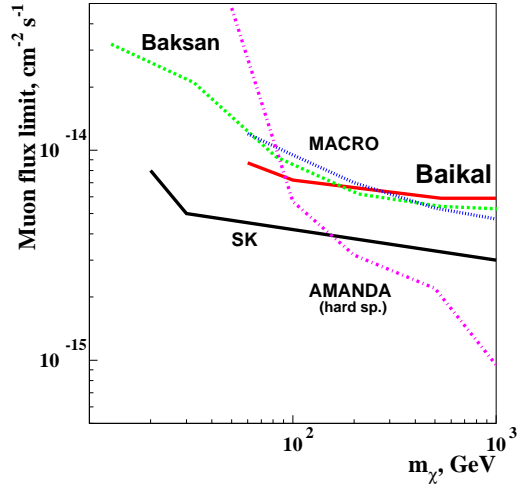}
\caption{Limits on the excess muon flux from the center of 
the Earth as a function of WIMP mass.
}
\label{fig4}
\end{figure}

Regarding the 24 detected events as being induced by atmospheric 
neutrinos, one can derive an upper limit on the additional flux of muons 
from the center of the Earth   
due to annihilation of neutralinos - the favored candidate for
cold dark matter.
The 90\% C.L. muon flux limits for six cones 
around the opposite zenith as well as muon flux limits for 
different neutralino masses
obtained with NT200 ($E_{\mbox{thr}}>$10 GeV)
in 1998/99 are shown in Fig. \ref{fig3} 
and Fig. \ref{fig4}, and
compared to limits obtained by 
Baksan \cite{BAKSANWIMP}, MACRO \cite{MACROWIMP}, 
Super-Kamiokande \cite{SKWIMP} and AMANDA 
(from the hard neutralino annihilation channels)
\cite{AMANDAWIMP}.

\vspace{-0.3cm}
\subsection{A search for fast magnetic monopoles}
\vspace{-0.3cm}
Fast magnetic monopoles with Dirac charge $g=68.5 e$ are 
interesting objects to search for with deep underwater neutrino telescopes.
The intensity of monopole Cherenkov radiation is $\approx$ 8300
times higher than that of muons. Optical modules of the 
Baikal experiment can detect such an object from a distance up 
to hundred meters.  
The processing chain for fast monopoles starts with the selection of
events with a high multiplicity of hit channels: $N_{hit}>30$. 
In order to reduce  the background  from downward atmospheric muons 
we restrict ourself to monopoles coming from the lower hemisphere. 
For an upward going particle the times of hit channels increase 
with rising z-coordinates from bottom to top of the detector. 
To suppress downward  moving particles, a cut on the value of the 
time--z--correlation, $C_{tz}$, is applied:             
\begin{equation}
  \label{eq1}
  C_{tz}= \frac{ \sum_{i=1}^{N_{hit}}(t_i- \overline t )(z_i- \overline
    z)} {N_{hit} \sigma_t \sigma_z}>0
\end{equation}
where $t_i$ and $z_i$ are time and z-coordinate of a fired channel, 
$\overline t$ and $\overline z$ are mean values for times and
z-coordinates of the event and $ \sigma_t$ and $ \sigma_z$ the 
rms--errors for time and z-coordinates.

\begin{figure}[htb]
\includegraphics[width=6.5cm,height=6.5cm]{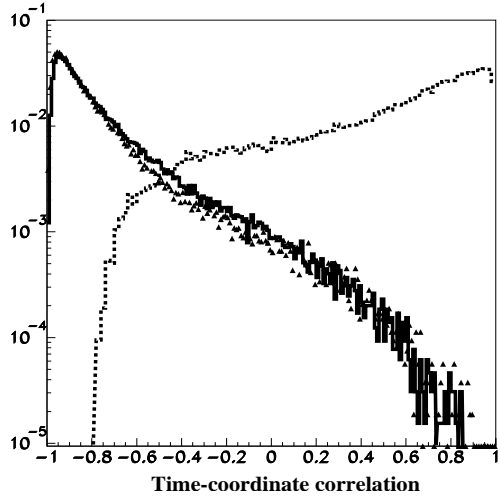}
\caption{
C$_{tz}$ distributions 
for experimental events (triangles),
simulated atmospheric muon events (solid), 
and simulated
upward moving 
relativistic magnetic monopoles (dotted);
multiplicity cut  $N_{hit}>30$.
}
\label{fig5}
\end{figure}

In Fig.~\ref{fig5}  we 
compare the $C_{tz}$-distribution for experimental data (triangles), simulated 
atmospheric muon events (solid curve)
with simulated upward moving monopole events
(dotted curve). 

Within 994 days 
of 
live time using in this analysis, about $ 3 \cdot 10^8$ events with $N_{hit} > 4 $
have been recorded, with 20943 of them satisfying cut 0 ($N_{hit}>30$ and 
$C_{tz}>0$). 
For further background suppression 
(see \cite{Mono05} for details of the analysis)
we use additional cuts, which essentially reject muon events 
and at the same time only slightly reduce
the effective area for relativistic monopoles
\footnote{Different values of cuts correspond to different
NT200 operation configurations.}:
\begin{enumerate}
\item  $N_{hit}>35$ and $C_{tz}>0.4 \div 0.6$
\item  $\chi ^2$ determined from reconstruction  has to be smaller than 3  
\item  Reconstructed zenith angle $\theta >100 \deg$
\item  Reconstructed track distance from NT200 center $R_{rec} >20 \div 25 $ m.
\end{enumerate}

No events from the experimental sample pass cuts 1-4.
The acceptances $ A_{eff} $ for monopoles with $ \beta=$1, 0.9 and 0.8  have been 
calculated 
for all NT200 operation configurations (various sets of operating channels).
For the time periods included, $ A_{eff} $ varies between
$3 \cdot 10^8$ and $6 \cdot 10^8 $cm$^2$sr (for $ \beta=1$).
From the non-observation of candidate events in NT200 and  the earlier
stage telescopes NT36 and NT96 \cite{baikal},  
a combined upper limit on the flux of fast monopoles 
with $90\%$ C.L. is obtained.   

\begin{figure}[htb]
\includegraphics[width=7cm,height=7cm]{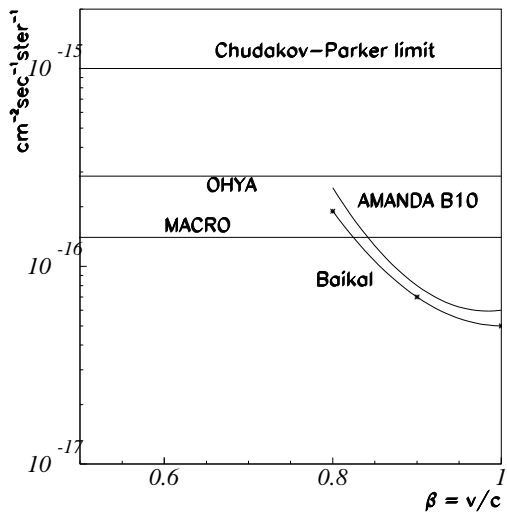}
\caption{
Upper limits on the flux of fast monopoles
obtained in this analysis (Baikal) and in other 
experiments.
}
\label{fig6}
\end{figure}

In Fig.~\ref{fig6} we compare 
this upper limit 
for an isotropic flux of fast monopoles 
obtained with the Baikal neutrino telescope  
to the limits from the underground 
experiments Ohya \cite{ohya}
and MACRO\cite{macro} and to the 
limit reported for the underice 
detector AMANDA\cite {AMANDAMON}.        
The Baikal limit is currently the most stringent one.

\vspace{-0.3cm}
\subsection{A search for extraterrestrial high-energy neutrinos}
\vspace{-0.3cm}
The BAIKAL survey for high energy neutrinos searches
for bright cascades produced at the neutrino interaction
vertex in a large volume around the neutrino telescope \cite{HEB}.
We select events with high multiplicity of hit channels $N_{\mbox{\small hit}}$,
corresponding to bright cascades. 
To separate high-energy neutrino events
from background events a cut
to select events with upward moving light signals has been developed.
We define for each event
$t_{\mbox{\footnotesize min}}=\mbox{min}(t_i-t_j)$,
where $t_i, \, t_j$ are the arrival times at channels $i,j$ 
on each string, and the minimum over all strings is calculated.
Positive and negative values of $t_{\mbox{\footnotesize min}}$ correspond to 
upward and downward propagation of light, respectively. 

\begin{figure*}
\includegraphics*[width=.5\textwidth,height=7.8cm]{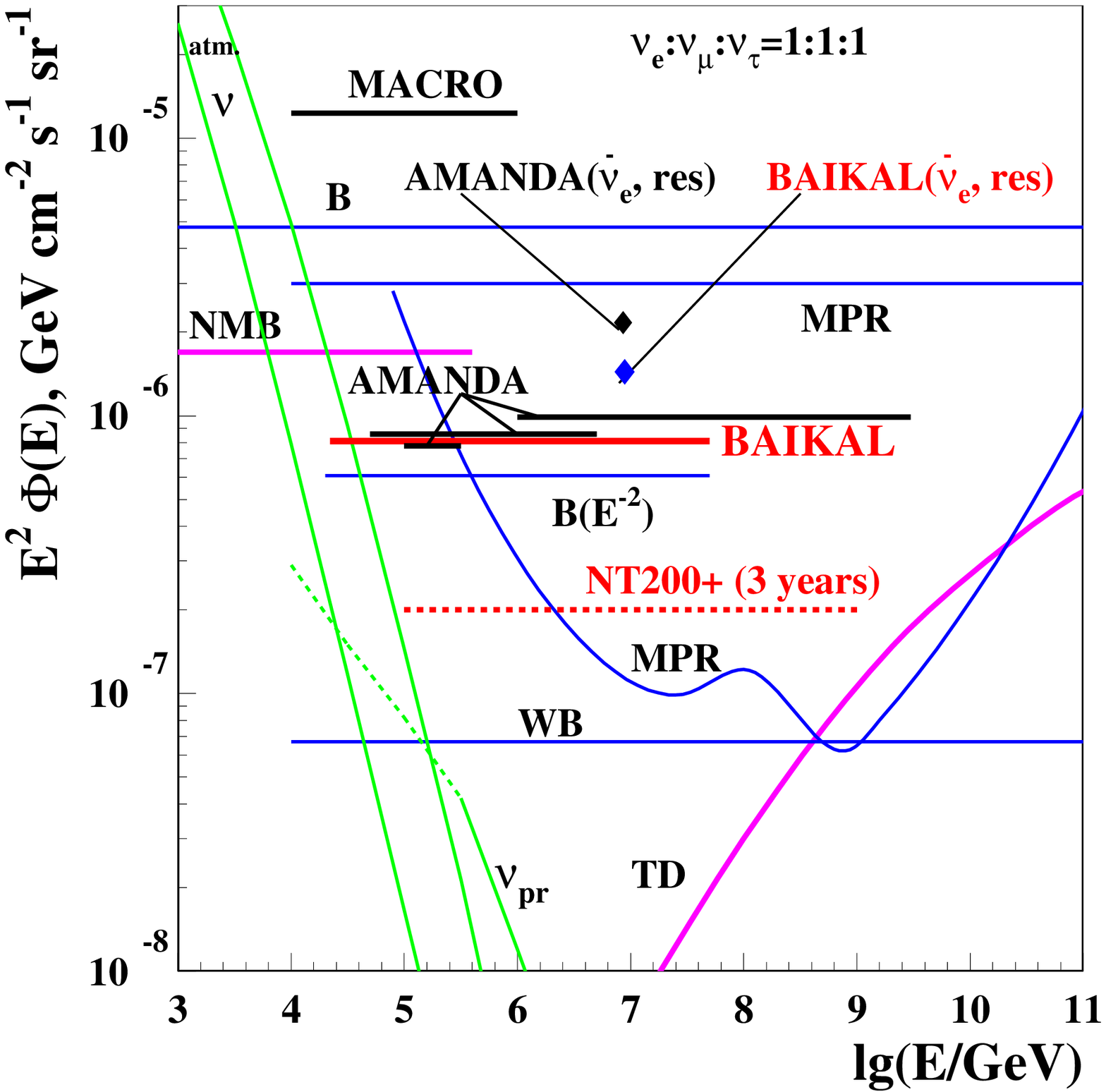}
\hfill
\includegraphics*[width=.55\textwidth,height=7.8cm]{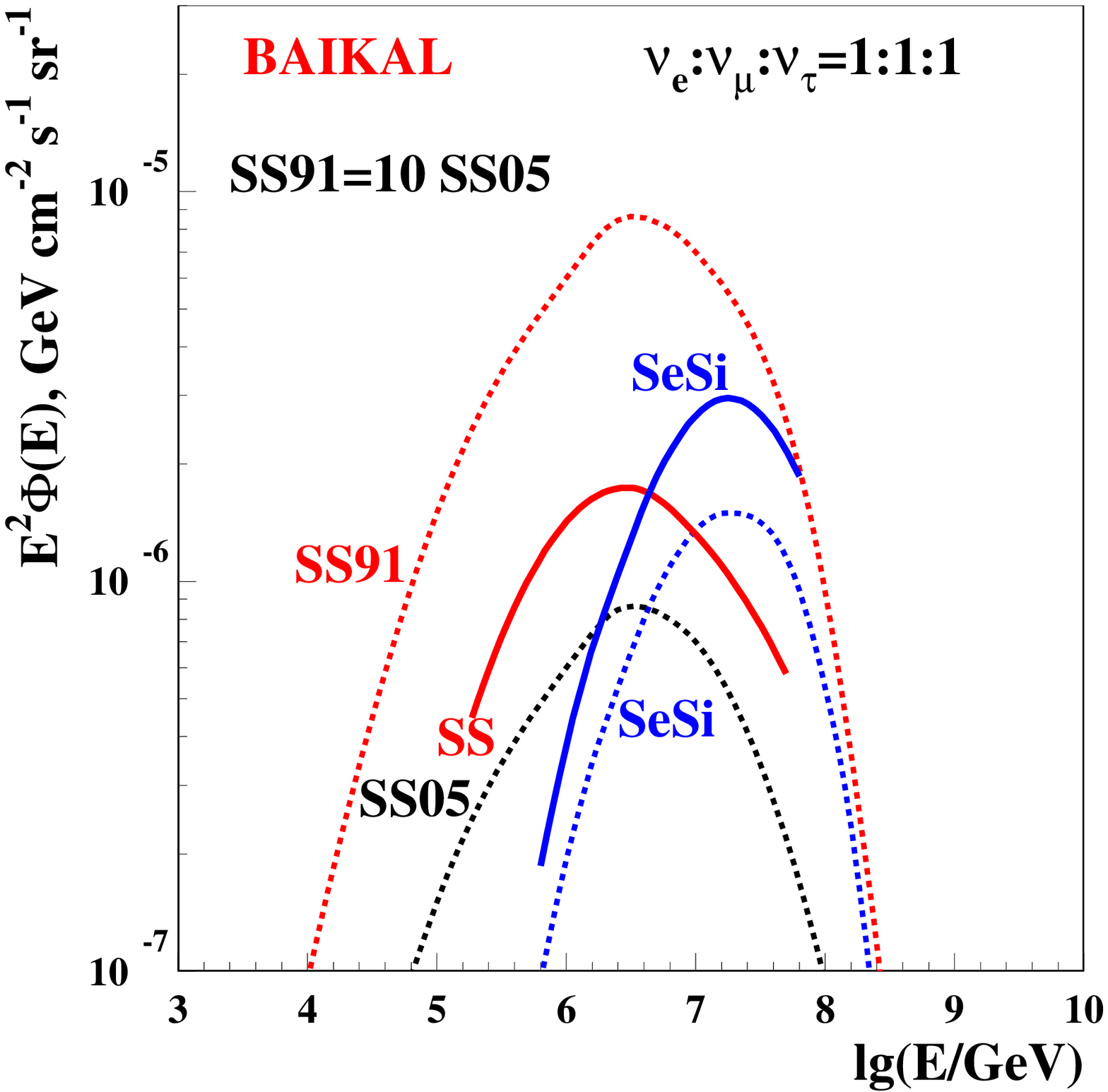}
\\
\caption{Left panel: all-flavor neutrino flux 
predictions in different models of neutrino sources 
compared to experimental upper limits to $E^{-2}$ fluxes obtained
by this analysis and other experiments (see text).
Also shown is the sensitivity expected for 3 live years of the
new telescope NT200+ \cite{RW,NT+2}.
Right panel: Baikal
experimental limits compared to two
model predictions. Dotted curves: predictions
from model SS \cite{SS}, SeSi \cite{SeSi} and SS05 \cite{SS05}. Full curves:
upper limits to spectra of the same shape.
Model SS is excluded (MRF=0.25), model SeSi is not
(MRF=2.12).
}
\label{fig7}
\end{figure*}

Within the 1038 days of the detector live time
between April 1998 and February 2003, 
$3.45 \times 10^8$ events with $N_{\mbox{\small hit}} \ge 4$ have been recorded. 
For this analysis we used 22597 events with hit channel multiplicity
$N_{\mbox{\small hit}}>$15 and  
$t_{\mbox{\footnotesize min}}>$-10 ns.
We conclude that 
data are consistent with simulated background
for both $t_{\mbox{\footnotesize min}}$ and $N_{\mbox{\small hit}}$ 
distributions. No statistically significant excess above the background 
from atmospheric muons has been observed. 
To maximize the sensitivity to a neutrino signal we introduce a cut in the 
($t_{\mbox{\footnotesize min}},N_{\mbox{\small hit}}$) phase space.

\begin{table}[htb]
\caption{Expected number of events $N_{\mbox{\footnotesize model}}$ and model rejection factors for models of
astrophysical neutrino sources.} 
\label{tab1}
  \begin{tabular}{@{}lcc|c}
\hline
 & \multicolumn{2}{c|}{BAIKAL} & AMANDA \cite{AMANDAHE,AMANDAMU,AMANDAMU2}\\
\hline
Model & $\nu_e+\nu_{\mu}+\nu_{\tau}$ & $n_{90\%}/N_{\mbox{\footnotesize model}}$ & 
$n_{90\%}/N_{\mbox{\footnotesize model}}$  \\
\hline
  10$^{-6}\times E^{-2}$ & 3.08 & 0.81 & 0.86  \\
  SS Quasar \cite{SS} & 10.00 & 0.25 & 0.21  \\
  SP u  \cite{SP}& 40.18 & 0.062 & 0.054  \\
  SP l \cite{SP}& 6.75 & 0.37 & 0.28  \\
  P $p\gamma$ \cite{P}& 2.19 & 1.14 & 1.99  \\
  M $pp+p\gamma$ \cite{M} & 0.86 & 2.86 & 1.19  \\
  MPR \cite{MPR}& 0.63 & 4.0 & 4.41  \\
  SeSi \cite{SeSi} & 1.18 & 2.12 & -  \\
  \hline
\end{tabular} 
\end{table}

Since no events have been observed which pass the final cuts
upper limits on the diffuse flux of extraterrestrial 
neutrinos are calculated. For a 90\% confidence level an upper limit 
on the number of signal events of $n_{90\%}=$2.5  is obtained 
assuming an uncertainty in signal detection of 24\% 
and a background of zero events.

A model of astrophysical neutrino sources, for which the total number
of expected events, $N_{\mbox{\footnotesize model}}$, is large than 
$n_{90\%}$, is ruled out at 90\% CL. 
Table \ref{tab1} represents event rates and model rejection factors (MRF) 
$n_{90\%}/N_{\mbox{\footnotesize model}}$ 
for models of astrophysical neutrino sources 
obtained from our search, as well as 
model rejection factors obtained recently
by the AMANDA collaboration \cite{AMANDAHE,AMANDAMU,AMANDAMU2}.

For an $E^{-2}$ behaviour of the neutrino spectrum and a flavor ratio 
$\nu_e:\nu_{\mu}:\nu_{\tau}=1:1:1$, the 90\% C.L. upper limit on the 
neutrino flux of all flavors obtained with the Baikal neutrino telescope  
NT200 (1038 days) is:
\begin{equation}
E^2\Phi<8.1 \times 10^{-7} 
\mbox{cm}^{-2}\mbox{s}^{-1}\mbox{sr}^{-1}\mbox{GeV}.
\label{eq2}
\end{equation}

For the resonant process 
with the resonant neutrino energy  
$E_0=6.3\times 10^6 \,$GeV 
the model-independent limit on $\bar{\nu_e}$ is: 
\begin{equation}
\Phi_{\bar{\nu_e}} < 3.3 \times 10^{-20}
\mbox{cm}^{-2}\mbox{s}^{-1}\mbox{sr}^{-1}\mbox{GeV}^{-1}.
\label{eq3}
\end{equation}

Fig. \ref{fig7} (left panel) shows our upper limit on 
the all flavor $E^{-2}$ diffuse flux (\ref{eq2})
as well as the model independent limit on the resonant $\bar{\nu}_e$ flux 
(diamond) (\ref{eq3}). Also shown are the limits obtained by AMANDA 
\cite{AMANDAHE,AMANDAMU,AMANDAMU2}
and MACRO \cite{MACROHE}, theoretical bounds obtained by 
Berezinsky (model independent (B) \cite{Ber3} and for an $E^{-2}$ shape
of the neutrino spectrum (B($E^{-2}$)) 
\cite{Ber4}, by Waxman and Bahcall (WB) \cite{WB1}, by Mannheim et al.(MPR) 
\cite{MPR}, predictions for neutrino fluxes from topological defects (TD) 
\cite{SeSi}, prediction on diffuse flux from AGNs according to Nellen et al. 
(NMB) \cite{NMB}, 
as well as the atmospheric 
conventional neutrino \mbox{fluxes \cite{VOL}} from horizontal and vertical 
directions ( ($\nu$) upper and lower curves, respectively) and atmospheric prompt 
neutrino fluxes ($\nu_{pr}$) obtained by Volkova et al. \cite{VPPROMPT}.
The right panel of Fig. \ref{fig7} shows our upper limits (solid curves) on 
diffuse fluxes from AGNs shaped according to the model of Stecker and 
Salamon (SS, SS05) \cite{SS,SS05} and of Semikoz and Sigl (SeSi) \cite{SeSi}, 
according to Table \ref{tab1}.

\vspace{-0.5cm}
\section{Conclusion}
The Baikal neutrino telescope NT200 is 
taking data since April 1998. 
The upper limit obtained for a diffuse 
($\nu_e+\nu_{\mu}+\nu_{\tau}$) 
flux with $E^{-2}$ shape
is $E^2 \Phi = 8.1 \times 10^{-7}$cm$^{-2}$s$^{-1}$sr$^{-1}$GeV.
The limits on fast magnetic monopoles and on  
a $\bar{\nu_e}$ flux
at the resonant  energy 6.3$\times$10$^6$GeV are presently the most
stringent.
To extend the search for diffuse extraterrestrial neutrinos 
with higher sensitivity, NT200 was significantly upgraded to NT200$+$,
a detector with about 5 Mton enclosed volume, 
which takes data since April 2005 \cite{RW,NT+2}.
The three-year sensitivity 
of NT200$+$ to the all-flavor neutrino flux
is approximately 
$2\times10^{-7}$cm$^{-2}$s$^{-1}$sr$^{-1}$GeV
for $E>$10$^2$ TeV (shown in Fig. \ref{fig7}).
In parallel with exploiting NT200$+$
we started research \& development activities towards
a Gigaton Volume Detector in Lake Baikal. 

\vspace{-0.5cm}
\ack
 This work was supported by the Russian Ministry of Education and Science, the 
  German Ministry of Education and Research and the Russian Fund of Basic Research
  (grants 05-02-17476, 04-02-17289, 04-02-16171, 05-02-31021, 05-02-16593), 
and by the Grant of  the President of Russia NSh-1828.2003.2.


\end{document}